\documentclass[fleqn,10pt]{wlscirep}
\usepackage[utf8]{inputenc}
\usepackage[T1]{fontenc}

\title{High-resolution European daily soil moisture derived with machine learning (2003-2020)}

\author[1,*]{Sungmin O}
\author[2]{Rene Orth}
\author[2]{Ulrich Weber}
\author[1,3,4]{Seon Ki Park}
\affil[1]{Department of Climate \& Energy System Engineering, Ewha Womans University, Korea}
\affil[2]{Department of Biogeochemical Integration, Max Planck Institute for Biogeochemistry, Jena, Germany}
\affil[3]{Center for Climate/Environment Change Prediction Research, Ewha Womans University, Korea}
\affil[4]{Severe Storm Research Center, Ewha Womans University, Korea}

\affil[*]{corresponding author(s): Sungmin O (sungmin.o@ewha.ac.kr)}

\begin{abstract}
{Machine learning (ML) has emerged as a novel tool for generating large-scale land surface data in recent years. ML can learn the relationship between input and target, e.g. meteorological variables and in-situ soil moisture, and then estimate soil moisture across space and time, independently of prior physics-based knowledge. Here we develop a high-resolution (0.1$^{\circ}$) daily soil moisture dataset in Europe (SoMo.ml-EU) using Long Short-Term Memory trained with in-situ measurements. The resulting dataset covers three vertical layers and the period 2003–2020. Compared to its previous version with a lower spatial resolution (0.25$^{\circ}$), it shows a closer agreement with independent in-situ data in terms of temporal variation, demonstrating the enhanced usefulness of in-situ observations when processed jointly with high-resolution meteorological data. Regional comparison with other gridded datasets also demonstrates the ability of SoMo.ml-EU in describing the variability of soil moisture, including drought conditions. As a result, our new dataset will benefit regional studies requiring high-resolution observation-based soil moisture, such as hydrological and agricultural analyses. The SoMo.ml-EU is available at figshare.}
\end{abstract}
\begin{document}

\flushbottom
\maketitle

\thispagestyle{empty}
\textcolor{blue}{\textit{This manuscript is a preprint currently under review.}}
\section*{Background \& Summary}
Because soil moisture is a key variable in water, energy, and biogeochemical cycles \cite{seneviratne_investigating_2010}, the availability of large-scale, high-resolution soil moisture datasets can facilitate diverse weather and climate applications, including monitoring and forecasting of hydrological and ecological extreme events \cite{bolten_evaluating_2010, orth_using_2014, wanders_suitability_2014, martinez-fernandez_satellite_2016, o_observational_2020, kroll_spatially_2022}. Soil moisture exhibits complex spatial and temporal dynamics, which undermines the usefulness of the direct use of sparse in-situ measurements for estimations or even predictions across larger areas \cite{famiglietti_field_2008, brocca_soil_2017}. Consequently, different approaches have been explored as a way of obtaining improved spatial data coverage. Physics-based land surface models have played the main role in providing continental-to-global scale soil moisture datasets \cite{rodell_2004, naz_3_2020, munoz-sabater_era5-land_2021}. Despite the fact that soil moisture values are not comparable and transferable among models \cite{koster_nature_2009}, the models can offer temporally and spatially continuous data across multiple soil layers. Satellite observations are another important resource, providing soil moisture estimates across large areas\cite{petropoulos_surface_2015, dorigo_esa_2017, chan_development_2018, yao_long_2021}, although their data are often missing due to the local retrieval conditions (e.g. presence of clouds or dense vegetation) or satellite revisit times. Moving beyond model and satellite-based datasets, novel machine learning (ML) approaches have been increasingly employed in recent years to generate large-scale soil moisture datasets, for instance, by filling the temporal and spatial gaps in satellite observations or by integrating multiple data sources \cite{park_downscaling_2017, mao_gap_2019, guevara_gap-free_2021}. \\[0.1in]
\indent
In this context, our previous study \cite{o_global_2021} (hereafter O21) used ML to generate a global soil moisture dataset — SoMo.ml — by upscaling in-situ measurements. We trained ML with meteorological forcing and soil moisture, such that the ML could learn emerging input-output relationships and thereby establish its own knowledge about processes. In this way, ML provides an attractive avenue to provide data that are independent from traditional modeling or satellite-based datasets. This is particularly useful for the regions where established approaches can not provide reliable estimates, e.g. high latitudes or densely-vegetated areas, because with another independent soil moisture data source, we can mitigate data uncertainty by comparing or using an ensemble of diverse datasets generated by different approaches \cite{guo_improving_2007, bai_performance_2018}. Further, ML-driven datasets can foster scientific discovery in the Earth and climate domain dominated by physics-based models, particularly for processes that are not sufficiently well represented in the models \cite{reichstein_deep_2019, geer_learning_2021}. For instance, SoMo.ml has contributed to observation-based analysis of drought-related ecosystem damages or heat events \cite{zhang_exacerbated_2021, bastos_vulnerability_2021, O_2022}. \\[0.1in]
\indent
Despite the availability of various data sources, acquiring high-resolution soil moisture data remains a challenge \cite{meng_high-resolution_2019, peng_roadmap_2021}. Various downscaling approaches have been developed to improve the spatial resolution of existing data \cite{sabaghy_spatially_2018}, however, the publicly available large-scale datasets are mostly available at coarse resolutions (25-50 km). To better reflect the land surface heterogeneity associated with soil and vegetation characteristics and, thereby, improve understanding of soil moisture heterogeneity and related hydrological and meteorological processes, there is a need for a more detailed picture of soil moisture dynamics across a broad range of relevant communities \cite{nayak_high-resolution_2018, vergopolan_field-scale_2021, abbaszadeh_high-resolution_2021, peng_roadmap_2021}. \\[0.1in]
\indent
Given the wide range of applications for soil moisture data, and the increasing need for high-resolution data, we aim to provide an updated ML-driven soil moisture dataset for Europe and document the added value of increased spatial resolution. To do this, we largely follow the methodology of O21; Long Short-Term Memory (LSTM) networks ingest meteorological forcings and static features as inputs and return estimated soil moisture as outputs (targets). LSTM is a special kind of recurrent neural network \cite{hochreiter_long_1997}, designed specifically to overcome the long-range temporal dependency problem of the traditional recurrent neural networks. This makes LSTM networks well suited to model soil moisture which, as a state variable and given the soil water holding capacity, integrates the meteorological forcing and thereby is driven by both concurrent and preceding meteorology\cite{o_global_2021, gao_modeling_2021, li_attention-aware_2022}. We scale point-level in-situ data to grid-scale target data by adjusting the mean and standard deviation of the in-situ data to those of ERA5-Land reanalysis 0.1$^{\circ}$ gridded soil moisture \cite{munoz-sabater_era5-land_2021} within the overlapping time period. In other words, the long-term mean and variation of the target data largely follow ERA5-Land, while daily variation is retrieved from in-situ measurements. As a result, our new dataset SoMo.ml-EU provides multi-layer daily soil moisture (0-10, 10-30, and 30-50 cm depths) for Europe at a 0.1$^{\circ}$ resolution. The resulting data show comparable performance to other widely-used gridded datasets. The most distinct feature of SoMo.ml-EU is that, by design, it closely follows daily variations in in-situ data, even more so than the downscaled previous version. Consequently, SoMo.ml-EU will be useful for various studies and applications that require high-resolution, observation-based soil moisture information. 

\section*{Methods}

\subsection*{Training data preparation}
\subsubsection*{Target In-situ soil moisture}
\begin{figure}[ht]
\centering
\includegraphics[width=\linewidth]{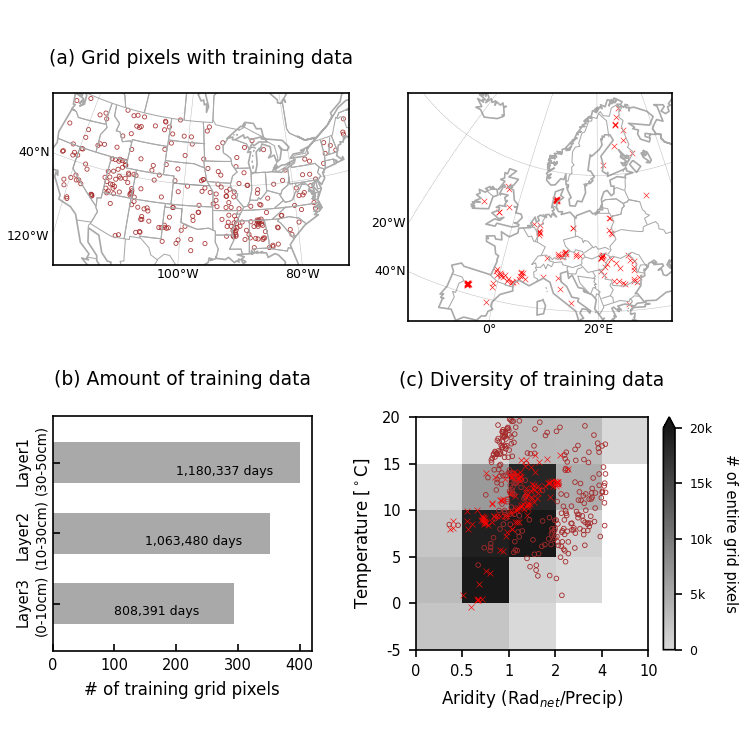}
\caption{(a) Spatial distribution of training data; brown circles and red markers show the grid pixels where in-situ soil moisture data are obtained in US and Europe, respectively. (b) The number of grid pixels and the total amount of training data for each layer. (c) Hydroclimatic diversity of origins of training data for the first layer; markers show the training data distribution across different climatic regimes defined by the long-term mean aridity and temperature of each grid pixel. Grey-scale colours represent the number of grid pixels for each regime across the European domain that SoMo.ml-EU covers. The same information for the second and third layers can be found in Fig. S1.}
\label{fig:fig1}
\end{figure}

We obtain daily in-situ soil moisture data from the International Soil Moisture Network (ISMN). ISMN is an international data hosting facility that collects, harmonises, and shares in-situ soil moisture measurements available around the globe\cite{dorigo_international_2011, dorigo_international_2021}. We consider in-situ data not only from the European domain but also from the US. In this way, we are able to obtain a greater amount of training data, which also represents more diverse climate conditions, as shown in Fig. \ref{fig:fig1}. The LSTM-based model is trained using the training data from all available locations jointly; more data helps the model to accurately describe input-output relationships and more diverse data enables the model to represent the variation of these relationships across conditions. The resulting accurate relationships, and their variations across conditions, form the backbone of the LSTM-based estimation of soil moisture in different time periods and regions. While the capacity of ML models to extrapolate input-output relationships beyond known conditions is limited, they can transfer knowledge between space and time at the same time\cite{robustness_2020}; e.g. the models learn about the input-output relationships in less-sampled semi-arid sites with many training data collected during dry periods at humid sites. Table \ref{Tab:table1} shows a full list of the ISMN networks contributing to the training data. \\[0.1in]
\indent
The collected in-situ data are then scaled using the long-term mean and standard deviation of ERA-Land gridded soil moisture. This scaling is intended to eliminate systematic biases in soil moisture means and variabilities among different types of sensors and calibration techniques while maintaining the daily variations observed directly by in-situ measurements, consequently yielding upscaled soil moisture information at the target resolution of 0.1$^{\circ}$. It is worth noting that temporal variations in point-level data have a wider footprint than absolute soil moisture values \cite{mittelbach_new_2012}. If there are more than one scaled measurement time series for a target grid pixel, we take an average of them. See O21 for further details about the data preparation. 

\subsubsection*{Meteorological forcing variables and static features}
We use temperature, precipitation, net radiation, and skin temperature as meteorological forcing variables for the first layer. Each layer is computed with a separate LSTM, and for the deeper layers, we additionally use simulated soil moisture in the upper layer(s). These variables were found to be the most influential variables according to our previous analysis in O21, which quantified the contribution of the respective variables to the LSTM model performance. Other relevant variables, such as specific humidity or wind, only had a marginal effect on the performance of the LSTM; moreover, it remains challenging to obtain reliable high-resolution gap-free and observation-based data for these variables over large areas. The previous version of SoMo.ml used only ERA5 reanalysis data, whereas the meteorological data for this study are pulled from more diverse data sources (see Table \ref{Tab:table2}) in order to reduce the dependency of the resulting data product on ERA5. Forcing data with a sub-daily resolution are aggregated to a daily scale, and all data are available at 0.1$^{\circ}$ resolution such that no spatial aggregation is applied. \\[0.1in]
\indent
We also use static features such as climatological values (long-term mean precipitation and aridity), topography (means and standard deviations of sub-grid scale elevation values), vegetation (forest and short vegetation fractions), and soil properties (sand and clay fractions for surface and deep layers). These static inputs have proven to be particularly important in surface soil moisture simulations\cite{o_global_2021}. For the vegetation and soil data that have a higher native resolution, we upscale the data to 0.1$^{\circ}$ x 0.1$^{\circ}$ resolution to match the spatial resolution of the target data. A comprehensive description of the meteorological forcing and static datasets is given in Table \ref{Tab:table2}. 

\subsection*{Model training and application}
We employ a modified version of LSTM architecture, Entity-Aware LSTM \cite{kratzert_2019}, that can explicitly differentiate between dynamic and static inputs, i.e. meteorological forcing and static features in our case. The inputs are normalised by subtracting the mean and dividing by the standard deviation to stabilise and accelerate training \cite{lecun_efficient_2012}. We determine the optimal set of hyperparameter values, which govern the model structures and training processes, through k-fold cross-validation (with $k=5$). The final model has 128 hidden units in one LSTM layer with one dense layer, and the dropout rate is set to 0.4. Another important hyperparameter is the length of the input sequence (i.e. lookback), corresponding to the number of prior time steps of meteorological data, e.g. from $t-365$ to $t-1$, used to predict soil moisture at a time step $t$. The lookback is typically set to 365 days in daily hydrologic simulations with an assumption that the LSTM model can learn the dynamics of an entire preceding annual cycle\cite{gauch_rainfallrunoff_2021}. In order to better specify this hyperparameter, and to study soil moisture memory characteristics in LSTM, we conduct an additional experiment to assess the effect of input sequence length on model performance as part of the model validation process (see the following section). \\[0.1in]
\indent
Finally, we train the LSTM model using the entire training dataset. The trained model with all established relationships between the input variables and soil moisture across concurrent and previous times is then applied across the entire European grid pixels. We repeat the simulations five times and take the averages of the simulations to compute the final soil moisture given the random initialisation of LSTM weights (learnable parameters).

\section*{Data Records}
SoMo.ml-EU provides volumetric soil moisture over the domain of latitudes 12.0 $^{\circ}$W to 45.0 $^{\circ}$E and longitudes 36.0 $^{\circ}$N to 71.5 $^{\circ}$N. The data covers the period 2003--2020, and the spatiotemporal resolution is 0.1$^{\circ}$ and one day. The data files are freely available at figshare. An example file name is 'SoMo.ml-EU\_\textit{LAYER}\_\textit{YYYY}.nc', with \textit{LAYER} and \textit{YYYY} standing for soil moisture layer depth and year, respectively.

\section*{Technical Validation}
We report the suitability of LSTM-based model for soil moisture simulations, focusing on two aspects. First, we assess the impact of preceding meteorological information on modelling performance. As mentioned, it is a key characteristic of LSTM networks to consider the relationship of the target variable with both concurrent and preceding input variable estimates. Second, we evaluate the ability of LSTM to spatially extrapolate learned input-output relationships. This knowledge transfer between locations is critical to generating enhanced quality data in domains with limited training data.\\[0.1in]
\indent
Furthermore, we report the performance of the final data product of SoMo.ml-EU through intercomparison with multiple gridded datasets. We employ 1) SoMo.ml v1, the previous version of SoMo.ml with a lower spatial resolution (0.25$^{\circ}$), 2) ERA5-Land, a replay of the land component of the ERA5 reanalysis \cite{munoz-sabater_era5-land_2021}, 3) GLEAM satellite-based data; we choose v3.5.b, which is based on satellite data only and no reanalysis is used \cite{martens_gleam_2017}, and 4) CLM-DA, high-resolution soil moisture generated from the Community Land Model with data assimilation of ESA CCI satellite observations \cite{naz_3_2020}. Finally, we compare the datasets using independent in-situ soil moisture data obtained from the COSMOS-Europe network\cite{bogena_cosmos-europe_2022}. It is a newly introduced soil moisture network in Europe, which offers the opportunity for an independent and comparative evaluation as these measurements are not used in the derivation of any of the gridded products we compare here. Additionally, we compare the dataset at locations of around 300 grid pixels, selected from every 20 x 20 pixels segment over the entire domain, to overcome the limitation of the low climate diversity represented by the COSMOS-Europe data, using the average of the considered datasets as a reference.

\subsubsection*{Role of preceding meteorological information}
\begin{figure}[ht]
\centering
\includegraphics[width=\linewidth]{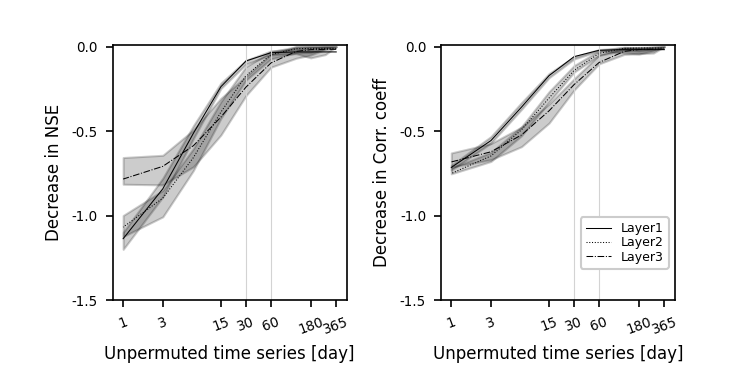}
\caption{Importance of preceding meteorological information for the performance of the resulting soil moisture simulation. This is expressed as performance reduction when perturbing part of the preceding meteorological information, as compared with the reference model performance using unchanged meteorological information. For example, results for 30 days are obtained with meteorological input data left unchanged for the last 30 days while being permuted for the first 365-30=335 days. Nash‐Sutcliffe efficiency and correlation coefficient are considered for performance evaluation. We repeat the simulations with different hyperparameters related to model architecture, e.g. the number of layers or hidden units. The shaded area shows the 0.2 to 0.8 quantiles of the metrics across model architectures. Note that x-axes use a logarithmic scale.}
\label{fig:fig2}
\end{figure}

First, we train the LSTM with input data covering the preceding 365 days and define its prediction performance as 'reference performance'. Next, we randomly permuted a part of the sequences except for the last $n$ time steps, i.e. from $t-365$ to $t-(n+1)$ days, where $t$ is the prediction time step. Consequently, only the later subsequent part of the time series, i.e. from $t-n$ to $t-1$ days, keeps its original information. Figure \ref{fig:fig2} shows the model performance for the surface soil moisture simulation using the partially permuted input sequences. The model performance significantly decreases when the unperturbed portion of the input time series becomes shorter than 30 to 60 days. On the other hand, the meteorological data prior to the preceding two months show a small contribution to the LSTM performance. This matches well with previous studies quantifying soil moisture memory time scales around several weeks to months, albeit with seasonal and regional variations \cite{koster_2001, orth_inferring_2013}. We repeat the simulations for the deeper layers and find a weaker model performance for the same degree of input data perturbation such that a similar model performance requires more preceding input information compared to the first layer simulations. This implies longer soil moisture memory in the deeper layers, which is consistent with the findings from analyses that used physics-based model data \cite{wu_time_2004}. \\[0.1in]
\indent
For the final model simulation, we keep the input sequence length (lookback) to 365 days to exploit the full potential of the meteorological information, i.e. giving the model the flexibility to account for potentially longer memory time scales occurring in specific regions and times. At the same time, these results have important implications. First, LSTM effectively extracts useful information from prior time steps and this explicitly contributes to more accurate soil moisture simulations. Therefore, a sufficient lookback period should always be considered. Second, we show the delayed effect of meteorological input on soil moisture and, therefore, the usefulness of LSTM to quantify soil moisture memory. Future research can focus on determining seasonal and regional variations of the soil moisture memory, potentially offering new insights into the conventional methods, e.g. using autocorrelation-based metrics \cite{orth_inferring_2013, mccoll_global_2017}.

\subsubsection*{Model performance across continents}
\begin{figure}[ht]
\centering
\includegraphics[width=\linewidth]{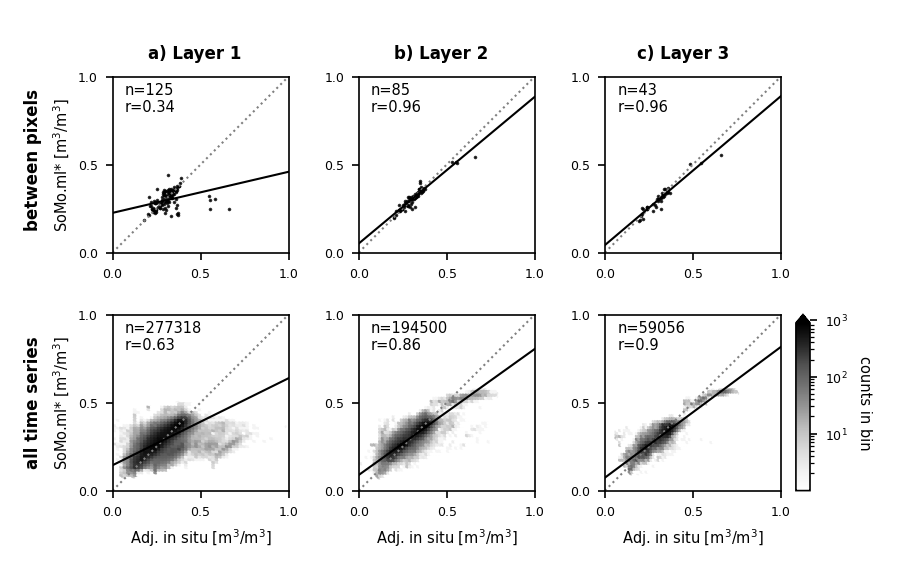}
\caption{Transfer learning capacity of the LSTM network. The model performance is validated in Europe after training with data from the US only. We compare (a) pixel-averaged soil moisture and (b) daily soil moisture from all grid pixels in Europe between target soil moisture and SoMo.ml* at each layer.}
\label{fig:fig3}
\end{figure}

We train the model with training data from the US domain only and validate the model performance over the European domain. In this way, we can evaluate the model with an emphasis on its capability to transfer knowledge across continents, and significantly minimise the risk of overfitting due to spatial autocorrelation between closely located training and validation data. The simulated soil moisture here is referred to as SoMo.ml* (Fig. \ref{fig:fig3}), because they are different from the final SoMo.ml-EU data generated from the model trained with the entire training data from the US and Europe both. We also report the model validation results from the five-fold cross-validation in the Supplementary material (Fig. S2).\\[0.1in]
\indent
Overall, the model shows satisfactory performance. The results clearly show how efficiently ML can transfer knowledge between locations, although relatively poorer performance is observed in Layer 1 (0-10 $cm$ depth), which is mainly due to outliers, i.e. target soil moisture from the five grid pixels with a very humid climate where the average soil moisture is above 0.5 $m^3/m^3$. This illustrates the main limitation of ML approaches. They can hardly extrapolate beyond the conditions covered by the training data. The US data are mostly sampled from arid and warm regions, so training data for humid and cold regions are lacking (see Fig. \ref{fig:fig1}(c)). The model shows a better performance for Layer 2 and Layer 3 (10-30 $cm$ and $30-50$ cm depths, respectively), which could be related to the lower temporal variability or to the relatively simple input-output relationships (deeper layer soil moisture is mainly determined by the upper layer soil moisture than dynamic meteorological information\cite{o_global_2021}) in the deeper layers that can be relatively easily reproducible by the model. Note that the model performance for wet regions is better ($r=0.8$ for correlation between pixels in Layer 1) according to the five-fold cross-validation in which the training data from Europe are included (Fig. S2). Therefore, the performance of the final SoMo.ml-EU over humid regions (similar to the training conditions) expected to be reasonable.  \\

\subsubsection*{Comparison against independent in-situ measurements}
\begin{figure}[ht]
\centering
\includegraphics[width=\linewidth]{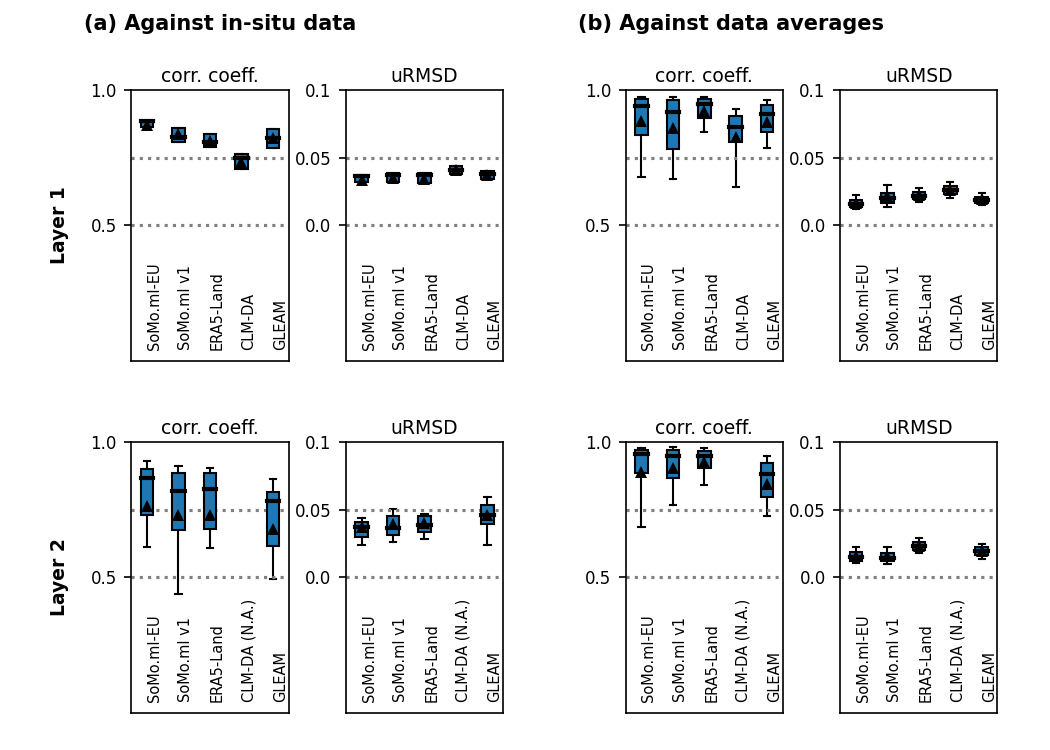}
\caption{(a) Comparison of soil moisture datasets against reference in-situ data over 6 and 22 grid pixels for Layers 1 and 2, respectively, and (b) against the average of all datasets at 287 randomly selected grid pixels from across the domain. Gridded soil moisture fields from SoMo.ml-EU, SoMo.ml v1, ERA-Land, CLM-DA, and GLEAM are considered. The climate diversity represented by the reference data used here is shown in Figs. S1 and S3. Correlation coefficient and unbiased root mean square difference are computed for the comparison. Triangles show means and box plot whiskers indicate the 0.2 to 0.8 quantiles of the metrics across grid pixels. See Fig. S4 for results for Layer 3.}
\label{fig:fig4}
\end{figure}

In Fig. \ref{fig:fig4}a, we compare SoMo.ml-EU, including multiple gridded datasets, against in-situ data obtained from the COSMOS-Europe network \cite{bogena_cosmos-europe_2022}. We select validation grid pixels that are not already included in our training data, although the COSMOS-Europe data only cover a relatively narrow range of climatic regimes. Note that COSMOS-EU does not include consistent data from all depths at all sites (see Fig. S1). The datasets with different native spatial resolutions are regridded to 0.1$^{\circ}$ using bilinear interpolation. We find that SoMo.ml-EU agrees closer with the in-situ data than the other gridded datasets including the previous SoMo.ml version, confirming the benefit of constructing high-resolution data rather than simply using interpolated data. \\[0.1in]
\indent
In terms of unbiased root mean square difference (uRMSD), SoMo.ml-EU shows smaller deviations from the in-situ data, while the median range of uRMSD across the datasets stays narrow between 0.04 to 0.05. SoMo.ml-EU and the other datasets all exhibit lower performance in Layer 2 than Layer 1, because the in-situ reference data are collected from more diverse climate conditions (see Fig. S1); reliable soil moisture derivations for extreme regimes are challenging to all approaches. In Fig. \ref{fig:fig4}b, the data are furthermore compared against the average of all the gridded datasets over the randomly selected grid pixels (Fig. S3) from which we do not have independent in-situ reference data. SoMo.ml-EU gives a reasonable performance within the range of the other datasets. Interestingly, ERA5-Land shows the highest agreement with the averages, which might be attributable to the dependency of the other datasets on ERA reanalyses. For instance, SoMo.ml used ERA5 meteorological data as the input and, for CLM-DA, ERA-Interim data is involved in the process of forcing data generation.\\

\subsubsection*{Comparison of soil moisture variability}
\begin{figure}[ht]
\centering
\includegraphics[width=\linewidth]{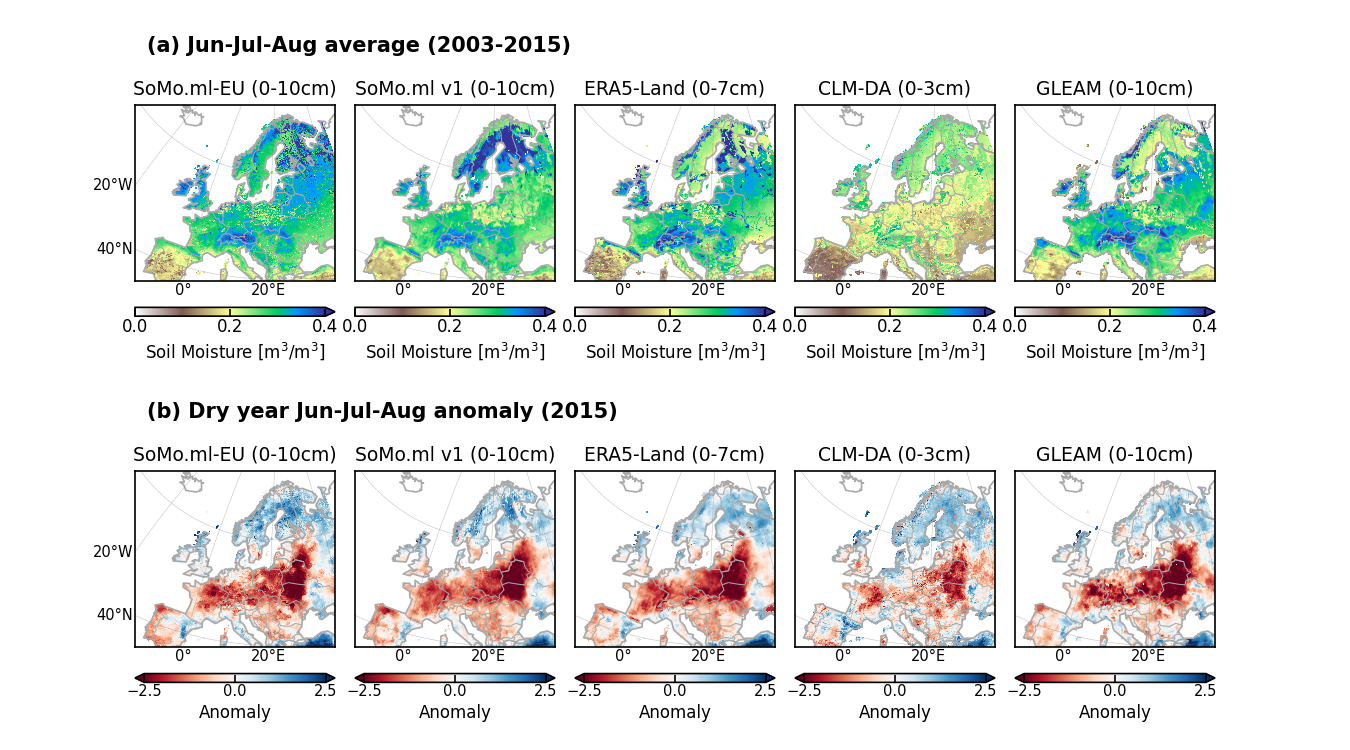}
\caption{Comparison of mean and extreme soil moisture from SoMo.ml-Eu with other state-of-the-art datasets; spatial distribution of (a) the average summer (JJA) soil moisture between 2003 and 2015 and (b) the normalised summer soil moisture anomaly in 2015. All datasets are regridded to 0.1 degrees. Results for Layers 2 and 3 are provided in Fig. S5.}
\label{fig:fig5}
\end{figure}

In this section, we assess the ability of SoMo.ml-EU to capture soil moisture variability, focusing on drought events. Figure \ref{fig:fig5}(a) presents the spatial distribution of the average surface soil moisture from SoMo-EU and the other gridded datasets during summer season (JJA). Overall, the spatial soil moisture patterns are similar across all datasets, with southwest to northeast gradients from dry to wet. Nonetheless, regional differences are observed, such as in mountain areas or high latitudes where observational data are typically lacking or highly uncertain due to, for instance, soil freezing. Therefore, high uncertainty is expected both for data-driven and observation-assimilated models. Even though we find good agreement, in terms of the spatial patterns, large deviations in the absolute values exist with the soil moisture from CLM-DA being especially generally underestimated. This dry bias can be due to a discrepancy in the definitions of the surface layer \cite{naz_3_2020}, e.g. 10 cm for SoMo.ml versus 3 cm for CLM-DA, or the use of observational data, e.g. in-situ measurements in SoMo.ml, while satellite data in CLM-DA. \\[0.1in]
\indent
We further compare the normalised soil moisture anomalies as depicted in Fig. \ref{fig:fig5}(b) during 2015 when large parts of Europe were affected by drought \cite{laaha_european_2017, ionita_european_2017}. The normalised anomalies are calculated by subtracting the long-term seasonal mean and then dividing by the standard deviation (z-score). The datasets show high agreement with similar patterns of positive and negative anomalies. The spatial extent of the affected area described by SoMo.ml-EU, i.e. pronounced negative anomalies over Central and Eastern Europe, matches well with the other datasets, except for the CLM-DA, which shows relatively weaker anomalies. We also compare seasonal anomalies over the entire overlapping years for different parts of Europe. The results show (Fig. S6) consistent temporal variations between SoMo.ml-EU and the other datasets, particularly for Southern Europe. On the other hand, for Northern Europe, we find relatively large deviations across the datasets, probably because of the high uncertainty in both observational data and physical model representation relating to freeze-thaw processes (ref). \\[0.1in]
\indent
In general, the soil moisture in SoMo.ml-EU shows similar spatial patterns and absolute values to those of ERA5-Land, which is no surprise given that ERA5-Land data are involved in the training data adjustment (Sect. 2.1). The interesting point here is, however, that despite the absence of training data for most of the grid pixels, SoMo.ml-EU shows reasonable spatial patterns and is in general agreement with the other datasets. This demonstrates how efficiently LSTM can spatially extrapolate target variables over large regions, which is enabled by our training strategy of using a combination of training data obtained from diverse climate regimes. This way, these results illustrate the usefulness of using in-situ soil moisture data from outside the European target domain to increase the amount and diversity of training data while benefiting from the transfer learning ability of the LSTM networks. \\

\section*{Usage Notes}

Despite the availability of large-scale soil moisture datasets from diverse sources, soil moisture information is mostly provided at coarse spatial resolutions (25-50 $km$), which limits the consideration of land surface heterogeneity and corresponding physical processes related to, for example, land-climate coupling. To meet the increasing requirement for high-resolution data \cite{peng_roadmap_2021, vergopolan_field-scale_2021}, we apply our distinctive approach, using ML trained with in-situ data, and deliver the high-resolution soil moisture over Europe. Consequently, SoMo.ml-EU offers daily soil moisture data at the spatial resolution of 0.1$^{\circ}$ over the period 2003-2020.\\[0.1in]
\indent
During the training-validation process, we confirm that ML can efficiently transfer knowledge from gauged to ungauged locations, even across continents. In our case, this is particularly useful for the simulations over arid regions because, in Europe, most in-situ data come from moderate climatic regions. Therefore, observational data covering more extreme conditions (e.g. arid regions) is only available from US networks. On the other hand, it remains challenging to obtain reliable soil moisture data for cold and humid regions, which could potentially introduce relatively high uncertainty in the final dataset in these regions. Over several years, more and more in-situ networks have been participating in the ISMN, and this study also utilises the in-situ data from the new network COSMOS-Europe. While those new data could be used in model training, we decided to keep them as independent data for validation purposes. Nonetheless, it is worth paying attention to publicly available big data because more diverse observational data is the key to reliable ML performance.\\[0.1in]
\indent
Our new dataset SoMo.ml-EU will benefit meteorological and hydrological applications that require higher-resolution soil moisture data, e.g. evaluation of up/downscaled products, analyses of regional drought events and the role of soil moisture for hydrological dynamics and land-climate interactions. The high temporal correlation between the SoMo.ml-EU and in-situ data is mainly due to the uniqueness of our approach in directly using in-situ measurements (i.e. daily anomalies) to inform the model. More importantly, we process the training data and operate the model at a higher spatial resolution than in the previous version, which better mirrors the spatial footprint of the in-situ measurements. Consequently, more efficient use is made of the information content of these measurements to the reproduce soil moisture dynamics over ungauged locations. The closer agreement with the in-situ data revealed in our specific evaluation does not necessarily mean that SoMo.ml-EU generally outperforms the other datasets, given the discrepancies among the datasets in terms of spatial scales or soil layer depths. Rather, it indicates the distinctive characteristic of our data. Consequently, SoMo.ml-EU can serve as a independent reference against which to evaluate other high-resolution datasets. At the same time thanks to its ML-based derivation it could be used jointly with other established datasets to form an ensemble of independent gridded state-of-the-art soil moisture datasets with potentially increased robustness in data sparse regions.

\section*{Acknowledgements}

This study is supported by Brain Pool program funded by the Ministry of Science and ICT through the National Research Foundation of Korea (NRF-2021H1D3A2A02040136). Rene Orth acknowledges funding from the German Research Foundation (Emmy Noether grant 391059971). We thank the ISMN (\url{https://ismn.geo.tuwien.ac.at/en/networks/}) and COSMOS-Europe (\url{https://doi.org/10.34731/x9s3-kr48}) participating networks for sharing their in-situ soil moisture data. 
\section*{Author contributions statement}

SO and RO designed the study. SO conducted the experiments. UW contributed to collecting the data. All authors contributed to the analysis of the results and to the writing of the manuscript.

\section*{Competing interests}

The authors declare that they have no conflict of interests.

\section*{Figures \& Tables}

\clearpage
\begin{table}[t]
\begin{center}
\begin{tabular}{|c|c|c|}
\hline
$Network$ & $Country$ & $Data~ aquisition$ \\
\hline
BIEBRZA-S-1\cite{BIEBRZA} & Poland & Layer1/2/3 \\
CALABRIA & Italy & Layer-/2/-/ \\
CAMPANIA\cite{brocca_soil_2011} & Italy& Layer-/2/-/\\
COSMOS\cite{zreda_cosmos_2012} & USA& Layer1/2/-/\\
FMI\cite{FMI}& Finland& Layer1/2/3 \\
FR-Aqui\cite{AQUI} & France & Layer1/2/3 \\
GROW\cite{GROW}& UK& Layer1/-/-/ \\
GTK & Finland & Layer1/2/3/\\
HOAL & Austria& Layer1/2/3/ \\
HOBE\cite{bircher_soil_2012} & Denmark & Layer1/2/-/ \\
HYDROL-NET-PERUGIA\cite{morbidelli_soil_2014} & Italy & Layer1/2/3/\\
IMA-CAN1\cite{biddoccu_long-term_2016} & Italy& Layer1/-/-/\\
IPE & Spain & Layer1/2/-/ \\
METEROBS & Italy & Layer1/2/3/ \\
MOL-RAO\cite{beyrich_site_2007} & Germany& Layer1/2/3/\\
REMEDHUS\cite{REMEDHUS} & Spain& Layer1/-/-/\\
RSMN & Romania& Layer1/-/-/\\
Ru-CFR & Russia & Layer1/2/3/\\
SCAN\cite{SCAN}& USA& Layer1/2/3/\\
SMOSMANIA\cite{calvet_situ_2007} & France& Layer1/2/-/\\
SWEX-POLAND\cite{marczewski_strategies_2010} & Poland& Layer1/2/-/\\
TERENO\cite{zacharias_network_2011} & Germany & Layer1/2/3/\\
UDC-SMOS\cite{schlenz_uncertainty_2012} & Germany& Layer1/2/3/\\
UMBRIA\cite{brocca_soil_2011} & Italy & Layer-/2/3/\\
UMSUOL & Italy& Layer1/2/3/\\
USCRN\cite{bell_us_2013} & USA & Layer1/2/3/ \\
VAS & Spain & Layer1/-/-/ \\
WEGENERNET\cite{kirchengast_wegenernet_2014} & Austria& Layer-/2/-/\\
WSMN\cite{WSMN} & UK & Layer1/-/-/\\
\hline
\end{tabular}
\caption{\label{Tab:table1}List of the ISMN networks considered to collect in-situ measurements.}
\end{center}
\end{table}

\clearpage
\begin{table}[ht]
\centering
\begin{tabular}{|p{1.3cm}|p{4.8cm}|p{2.5cm}|p{6cm}| }
\hline
& Variable & Source & Description\\
\hline
Dynamic & Air temperature & MSWX-Past\cite{beck_mswx_2022} & Bias-corrected and downscaled Climatic Research Unit Time Series data\\
& Precipitation &  MSWEP\cite{beck_mswep_2019} & Gauge, satellite, and reanalysis merged data\\
& Net surface radiation & ERA5-Land\cite{munoz-sabater_era5-land_2021} & ECMWF land reanalysis\\
& Land surface temperature & ERA5-Land\cite{munoz-sabater_era5-land_2021} & ECMWF land reanalysis\\
& Soil moisture from upper layer(s) & SoMo.ml-EU & ML-based soil moisture produced in this study\\
\hline
Static & Mean precipitation & ERA5-Land\cite{munoz-sabater_era5-land_2021} & Long-term mean precipitation\\
& Aridity & MSWEP\cite{beck_mswep_2019}, ERA5-Land\cite{munoz-sabater_era5-land_2021} & Ratio of net radiation (unit converted to $mm$) to precipitation\\
& Topography & ETOPO1\cite{amante_etopo1_2009} & Mean and standard deviation of sub-grid scale elevations\\
& Land cover & Harmonized World Soil Database v1.2\cite{harmonised_2008} & Forest and short vegetation fraction computed based on six geographic datasets\\
& Soil type & Harmonized World Soil Database v1.2 (regridded)\cite{wieder_regridded_2014} & Clay and sand fraction based on regional and national updates of soil information\\
\hline
\end{tabular}
\caption{\label{Tab:table2}Description of meteorological forcing variables and static features. }
\end{table}

\end{document}